\documentclass[10pt,a4]{article}
\newcommand*{\be}{\begin{eqnarray}}
\newcommand*{\ee}{\end{eqnarray}}
\newcommand*{\ket}[1]{\ensuremath{|#1\rangle}}
\usepackage{graphicx}%
\usepackage{pslatex}%
\usepackage[small]{caption}%
\begin{document}
% Please delete and replace with your own information accordingly.                                                %
\title{\bf{Exact Abelian and Non-Abelian Geometric Phases}}%

\author{ $^{\rm 1}$Chopin Soo, $^{\rm 2}$Huei-Chen Lin\\%
{\footnotesize{$${Department of Physics, National Cheng Kung University, Tainan 701, Taiwan}}}\\%
%{\footnotesize$^{2}${Faculty of Science,International Islamic University Malaysia, Jalan Bukit Istana 25250 Kuantan, Pahang, %Malaysia.}}\\%
{\footnotesize{\it{email: $^{\rm 1}$\underline{cpsoo@mail.ncku.edu.tw}, $^{\rm 2}$\underline{calc13@gmail.com}}}}}%
\date{}
\maketitle
\begin{abstract}
The existence of Hopf fibrations $S^{2N+1}/S^1 = CP^N$ and $S^{4K+3}/S^3 = HP^K$ allows us to treat the
Hilbert space of generic finite-dimensional quantum systems as the total bundle space with
respectively $U(1)$ and $SU(2)$ fibers and complex and quaternionic projective spaces as base
manifolds. This alternative method of studying quantum states and their evolution reveals the
intimate connection between generic quantum mechanical systems and geometrical objects. The
exact Abelian and non-Abelian geometric phases, and more generally the geometrical factors for
open paths, and their precise correspondence with geometric Kahler and hyper-Kahler
connections will be discussed. Explicit physical examples are used to verify and exemplify the
formalism.
\end{abstract}

%\noindent \footnotesize{\it {\bf Keywords:} Quantum Information$^{\rm 1}$,
%Asia-Pacific$^{\rm 2}$}
\stepcounter{section}
\section*{Introduction}\small
The study of geometric phases in quantum mechanics is a fruitful and active endeavor (see, for
instance, Ref. \cite{Resource} for a review, and references therein for a selected sample of the literature and
history of the subject). It reveals that fundamental geometrical structures are present in generic
quantum systems; and the rich and exact interplay that can exist between geometrical mathematical
structures (e.g. Hopf fibrations), physical solitons (monopoles and instantons, just to cite lowest dimensional
examples) and generic quantum systems is both fascinating and of great pedagogical
value.
\\
\\
The existence of Hopf fibrations $S^{2N+1}/S^1 = CP^N$ and $S^{4K+3}/S^3 = HP^K$  allows us to treat the Hilbert
space of generic finite-dimensional quantum systems as the total bundle space with respectively $U(1)$
and $SU(2)$ fibers and complex and quaternionic projective spaces as base manifolds. (In the latter case of quaternionic projective space,
only even dimensional systems are permitted). This pedagogical review relies heavily on two recent in-depth studies of Abelian and non-Abelian geometric phases in quantum mechanics\cite{Lin and Wang}. Previously Aharonov-Anandan presented the Abelian geometric phase as the difference between the total and dynamical phase\cite{AA}, and Page formulated the result in terms of complex projective Hopf fibrations\cite{Page}. Non-Abelian geometric phases were also discussed in terms of Hopf fibrations by Adler and Anandan\cite{AAdler}. A general discussion on Hopf fibrations can be found in \cite{Trautman}. The method of using Hopf fibrations to study quantum states and their evolution reveals the intimate connection between generic quantum mechanical systems and geometrical objects. The exact Abelian
and non-Abelian geometric phases, and more generally the geometrical factors for open paths, and their precise correspondence with geometric Kahler and hyper-Kahler connections will be discussed.
The emphasis here is on the applicability of the formulation to generic finite-dimensional systems and on the exactness of the resultant geometric phases. Explicit physical examples are used to verify and exemplify the formalism.

\stepcounter{section}
\section*{Hilbert space of finite-dimensional quantum systems and Hopf fibrations}\small
Consider the Hilbert space of an arbitrary ``$(N+1)$-state'' pure system: Let $\{|\alpha\rangle\}$, $\alpha=0,1,\cdots,N$ be a time-independent orthonormal basis. An arbitrary normalized state may be expressed as:
\be|\Psi\rangle=\frac{z^\alpha}{\sqrt{\bar{z}^\beta
z^\beta}}|\alpha\rangle\equiv c^\alpha|\alpha\rangle;
\ee
wherein $\vec{z}(t)=(z^0(t),z^1(t),\cdots,z^N(t))\in C^{N+1}-\{0\}$. Writing the complex coefficients $c^\alpha=x^\alpha+iy^\alpha$ as real numbers $x^\alpha\in R$ and  $y^\alpha\in R$, and noting the normalization condition, \be 1=\sum\limits^N_{\alpha=0}\left|c^\alpha\right|^2=\sum\limits^N_{\alpha=0}\left[(x^\alpha)^2+(y^\alpha)^2\right],\ee lead to the conclusion of the correspondence $\left\{c^\alpha\right\}\Leftrightarrow\left\{x^\alpha,
y^\alpha\right\}\in S^{2N+1}$. Thus the Hilbert spaces of respectively 2-state, 3-state, 4-state, 5-state, $\cdots$ systems are associated with
$S^3, S^5, S^7, S^9, \cdots,$ i.e. odd-dimensional spheres which have very special properties!
Besides the simple Hopf fibration $S^M/\left\{+1, -1\right\}=RP^M$ over M-dimensional real projective space and the
octonionic Hopf fibration $S^{8L+7}/S^7=OP^1=S^8$, the other two series of Hopf fibrations over complex and quaternionic projective spaces are of great interest. These are the  complex Hopf fibrations: $S^{2N+1}/S^1=CP^N$ over N-dimensional complex projective spaces (e.g. $S^3/S^1=[CP^1=S^2]$ (Dirac monopole)), and the
quaternionic Hopf fibrations: $S^{4K+3}/S^3=HP^K$ over K-dimensional quaternionic projective spaces
(e.g. $S^7/[S^3=SU(2)]= [HP^1=S^4]$ (BPST instanton)).

\stepcounter{section}
\section*{Complex Hopf fibration over $CP^N$ and exact Abelian geometric phase}\small
Complex projective spaces $CP^N$  are defined as spaces of $\vec{z}$ modulo the equivalence relation
$\vec{z}(t)\sim\lambda(t)\vec{z}(t); \lambda \in C-\{0\}$ i.e. with $(z^0, z^1, \cdots, z^N)$ and $(\lambda z^0, \lambda z^1, \cdots,
\lambda z^N)$ identified.
In any local patch or chart $U_{(\eta)}$ wherein $z^\eta\neq0$, the inhomogeneous coordinates
$\zeta^\alpha_{(\eta)}(t)=z^\alpha(t)/z^\eta(t)$ are well-defined, and we can pass from homogeneous coordinates $z^\alpha$ to
$\zeta^\alpha$ which is explicitly invariant under the complex $\lambda$ scaling. The Hopf projections for ${C}^{N+1}-\left\{0\right\}\rightarrow
S^{2N+1}\rightarrow CP^N$ can be explicitly realized by
\be z^\alpha\rightarrow
c^\alpha\equiv\frac{z^\alpha}{\sqrt{\bar{z}^\beta
z^\beta}}\rightarrow
c^\alpha/c^0=z^\alpha/z^0=\zeta^\alpha_{(\eta=0)};\ee
with each projection specified in local chart $U^{(\eta=0)}$ and extended to the atlas of all charts, $\cup U^{(\eta)}$.
This constitutes an explicit Hopf map of the $S^{2N+1}$ bundle over
$CP^N$ base manifold with $U(1)$ fiber.
\\
\\
The exact formula of geometric factor in quantum mechanics can be obtained from the following considerations:
Locally, $S^{2N+1}\sim\left\{{\rm part\, of\, } CP^N\right\}\times S^1$  and we may express $|\Psi\rangle$ in local coordinates.
In the local patch $U^{(0)}$ wherein $\zeta^\alpha=z^\alpha/z^0$,
and $e^{i\phi_{z^0}}\equiv z^0/|z^0|$  lead to
\be |\Psi\rangle=\frac{z^\alpha}{\sqrt{\bar{z}^\beta
z^\beta}}|\alpha\rangle =c^\alpha|\alpha\rangle
=e^{i\phi_{z^0}}\frac{\zeta^\alpha}{\sqrt{\bar{\zeta}^\beta
\zeta^\beta}}|\alpha\rangle.\ee

Substituting into the Schrodinger equation, $i\hbar\frac{d}{dt}|\Psi\rangle=H(t)|\Psi\rangle$ yields
\be\frac{d\phi_{z^0}}{dt}+\frac{\bar{\zeta}^\alpha (d\zeta^\alpha/dt)
- \zeta^\alpha(d\bar{\zeta}^\alpha/dt)}
       {2i\bar{\zeta}^\beta \zeta^\beta}
  =-\frac{\bar{\zeta}^\alpha H_{\alpha\beta} \zeta^\beta}{\hbar(\bar{\zeta}^\eta
  \zeta^\eta)}.\ee
Identifying $A \equiv \frac{\bar{\zeta}^\alpha(d\zeta^\alpha/dt)-\zeta^\alpha(d\bar{\zeta}^\alpha/dt)} {2i\bar{\zeta}^\beta\zeta^\beta}dt$ implies the overall phase can be solved as
\be \phi_{z^0}(t)=\phi_{z^0}(0)-(\int^{\zeta(t)}_{\zeta(0)}A)-\frac{1}{\hbar}\int^{t}_{0}\langle\Psi|H|\Psi\rangle
dt .\ee
It follows that the generic state is expressible, in terms of $\zeta$-coordinates of $CP^N$ and the phase $\phi_{z^0}$, as
\be|\Psi(t)\rangle=\frac{z^\alpha(t)}{\sqrt{\bar{z}^\beta(t)
z^\beta(t)}}|\alpha\rangle
=e^{i\phi_{z^0}(t)}\frac{\zeta^\alpha(t)}{\sqrt{\bar{\zeta}^\beta(t)
\zeta^\beta(t)}}|\alpha\rangle.\ee  Moreover, in the overlap $U_{(\eta)}\cap U_{(\xi)}$ , we have $\zeta^\alpha_{(\xi)}=z^\alpha/z^\xi=(z^\eta/z^\xi)\zeta^\alpha_{(\eta)}
\forall\, \alpha$, and the transition function $(z^\xi/z^\eta)\equiv
Re^{i\phi}\in C^1$.  The geometric connection is thus revealed to be \be A \equiv
-i\frac{{\bar \zeta}^\alpha\,d\zeta^\alpha - \zeta^\alpha
d{\bar\zeta}^\alpha}{2 {\bar\zeta}^\beta \zeta^\beta} =
-i\frac{{\bar \zeta}^i\,d\zeta^i - \zeta^i d{\bar\zeta}^i}{2 (1 +
{\bar\zeta}^j \zeta^j)}, \qquad  j =1,2,... N;
\ee
which is an Abelian  connection whose
curvature is $F = dA = 2K$, wherein $K$ is the Kahler
2-form (which is real and closed ($dK =0$)), while $CP^N$ which is a Kahler-Einstein manifold with the Fubini-Study metric.
\\
\\
The preceding formulas straightforwardly imply that the overlap function at different times is given by
\be \langle\Psi(T)\ket{\Psi(o)} &=&
 \frac{{\bar\zeta}^\alpha(T)\zeta^\alpha(o)}
       {[{\bar\zeta}^\beta(T)\zeta^\beta(T)]^{\frac{1}{2}}
[{\bar\zeta}^\kappa(o)\zeta^\kappa(o)]^{\frac{1}{2}}}e^{-i({\phi}_{z^0}(T)-{\phi}_{z^0}(o))}\cr
&=&\frac{{\bar\zeta}^\alpha(T)\zeta^\alpha(o)}{[{\bar\zeta}^\beta(T)\zeta^\beta(T)]^{\frac{1}{2}}
[{\bar\zeta}^\kappa(o)\zeta^\kappa(o)]^{\frac{1}{2}}}
 \exp\left(i\int^{\zeta(T)}_{\zeta(o)}A +\frac{i}{\hbar}\int^T_o\,\langle{\Psi(t)}|H(t)\ket{\Psi(t)}dt\right).
\ee
\\
 By subtracting $\int^T_o\,\langle{\Psi(t)}|H(t)\ket{\Psi(t)}dt$ which is referred to as the "dynamical phase", the geometric phase factor is the residual entity in the overlap function.  In the special case of a closed path $c=\partial S$ bounding a
two-surface $S$, $\zeta^\alpha(T) = \zeta^\alpha(o)$ ({\it closed
path means the wave function at $o$ and $T$ differs by only a total
phase} $\Longleftrightarrow \zeta^\alpha(T)= \zeta^\alpha(o) \,\forall\, \alpha$), the geometric phase factor with $\zeta_{(\eta
=0)}^\alpha= z^\alpha/z^0$; $\zeta^0_{(\eta=0)} = 1$ results in the geometric phase \be \arg[e^{i\oint_{c=\partial S} A}] =
\arg\big[\exp\big(\oint_c \frac{{\bar \zeta}^i\,d\zeta^i - \zeta^i
d{\bar\zeta}^i}{2 (1 + {\bar\zeta}^j \zeta^j)}\,\big)\big] = \int_S
F.\ee
\\
{\bf A note on the gauge symmetry of the geometric phase:}
\\
 There are {\it two} connections: $\langle{\Psi(t)}|\,{{d}\over{dt}}\ket{\Psi(t)}dt$, and the Kahler connection $A \equiv
-i\frac{{\bar \zeta}^\alpha\,d\zeta^\alpha - \zeta^\alpha
d{\bar\zeta}^\alpha}{2
{\bar\zeta}^\beta \zeta^\beta}$. They are related by
\be
 \langle{\Psi(t)}|\,{{d}}\ket{\Psi(t)} = A_{(\eta)} + d\phi_{(\eta)}(t);
 \ee wherein $\phi_{(\eta)} = z^{\eta}/|z^{\eta}|$. The L.H.S begets additional term  $d\chi(t)$ under $\ket{\Psi(t)} \mapsto e^{i\chi(t)}\ket{\Psi(t)}$. Consistently, on the R.H.S. $\phi_{(\eta)}(t) \mapsto \phi_{(\eta)}(t) + \chi(t)$.  But $A(\zeta)$ remains explicitly unchanged (an overall scaling for all $z^\alpha$ does not change $\zeta^\alpha \equiv z^\alpha/z^\eta$)! In other words, despite the similarities with the approach by Aharonov-Aharonov \cite{AA}, the Kahler potential $A$ does {\it not} gauge the symmetry $\ket{\Psi(t)} \mapsto e^{i\chi(t)}\ket{\Psi(t)}$ (which Aharonov-Anandan advocated \cite{AA}). Rather, the Kahler connection $A$ transforms as an Abelian $U(1)$ gauge potential under local coordinate transformations between patches and gauges this symmetry. In the overlap $U_{(\eta)}\cap U_{(\xi)}$, the coordinates are related by $\zeta^\alpha_{(\xi)}=z^\alpha/z^\xi=(z^\alpha/z^\xi)\zeta^\alpha_{(\eta)} \forall \alpha$; thus the transition function is just $(z^\xi/z^\eta)\equiv
 Re^{i\chi}\in C^1$. Under this change of coordinates, the
 connection $A=-i\frac{\bar{\zeta}^\alpha d\zeta^\alpha-\zeta^\alpha
 d\bar{\zeta}^\alpha}{2\bar{\zeta}^\beta\zeta^\beta}$ transforms as
 $A \mapsto A'=A+d\chi$. The geometric
 phase/factor and the state remain invariant under such coordinate transformations between different patches.

\stepcounter{section}
\section*{Explicit examples}\small
{\bf Generic qubit systems, $S^3/S^1=CP^1$, and  the Dirac monopole:}
\\
\\
A generic qubit (or 2-state) system corresponds to the Hopf fibration $S^3/[U(1)=S^1]=[CP^1=S^2]$.
For the qubit system, the state is
\be |\Psi\rangle= \sum^1_{\alpha =0} c^\alpha|\alpha\rangle
=\sum^1_{\alpha, \beta =0}e^{i\phi_{z^0}}\frac{\zeta^\alpha}{\sqrt{\bar{\zeta}^\beta
\zeta^\beta}}|\alpha\rangle.\ee
 The explicit parametrization $c^0 = e^{{i}(\chi-\frac{\phi}{2})}\cos(\frac{\theta}{2})$ and $c^1=
e^{{i}(\chi+\frac{\phi}{2})}\sin(\frac{\theta}{2})$ with $|c^0|^2 +
|c^1|^2=1$, leads to the Hopf map projection
\be c^\alpha(\chi,\theta,\phi) \rightarrow \zeta^\alpha \in
{CP^1} : \zeta^0 = c^0/c^0 =1,  \zeta^1= c^1/c^0 =
e^{i\phi}\tan(\frac{\theta}{2}).\ee
The geometric Kahler connection computed in accordance with our previous discussion is
\be A=\frac{{\bar\zeta}^1d\zeta^1-\zeta^1d{\bar\zeta}^1}{2i(1+{\bar\zeta}^1\zeta^1)}
= \frac{1}{2}(1-\cos\theta)d\phi,\,
e^{i\phi_{z^0}}=z^0/|z^0|=c^0/|c^0|= e^{i(\chi-\frac{\phi}{2})},\ee  which is precisely the gauge potential for a Dirac  monopole connection with Chern number \be\frac{1}{2\pi}\int F = \frac{1}{2\pi}\int dA =
\frac{1}{2\pi}\int^\pi_{\theta=0}\int^{2\pi}_0\,
\frac{1}{2}\sin{\theta}d\theta\wedge d\phi = 1.\ee
Note also that the local chart fails at the south pole $\theta =\pi$
where $c^0$ vanishes, and we need more than one patch for the
atlas. A chart which fails only at the north pole ($\theta=0$) is
$\zeta^\alpha_{(\eta=1)} = c^\alpha/c^1$. In the overlap
$U^{(0)}\bigcap U^{(1)}$, we have $\zeta^\alpha_{(0)} =
(c^1/c^0)\zeta^\alpha_{(1)} $ with transition function $(c^1/c^0)=
e^{i\phi}\tan(\frac{\theta}{2})$. Moreover, the phase of the
coordinate transition function $e^{i\phi}\tan(\frac{\theta}{2})$ is
precisely $\phi$ ; hence following our discussions in section VII,
$A_{(0)} = A_{(1)} + d\phi = A_{(1)} + e^{i\phi}i de^{-i\phi}$. The
monopole charge can also be deduced, via the Wu-Yang formulation,
from the $\Pi_{1}(U(1))$ homotopy map of the transition function,
$e^{i\phi}: \phi \in S^1 \rightarrow e^{i\phi} \in U(1)=S^1$, which
has winding number 1. Note the distinction between the Wu-Yang
transition function relating the monopole potentials $A_{(0)}$ and
$A_{(1)}$ (which are connected by gauge transformation $e^{i\phi}
\in U(1)$) and the transition function
$e^{i\phi}\tan(\frac{\theta}{2})$ between coordinate patches which
is a complex scaling. Remarkably the setup in the previous sections
yield these results self-consistently.
Furthermore, according to the rules of the formalism, the general state is
\be\ket{\Psi(t)}&=&\frac{z^\alpha(t)}{\sqrt{{\bar
z}^\beta(t)
z^\beta(t)}}\ket{\alpha}=e^{i\phi_{z^0}(t)}\frac{\zeta^\alpha(t)}{\sqrt{{\bar
\zeta}^\beta(t) \zeta^\beta(t)}}\ket{\alpha}\cr
   &=&\frac{e^{i\phi_{z^0}(t)}}{\sqrt{(1 + \tan^2({\theta}/{2})}}\big[ \ket{0}+ e^{i\phi}\tan({\theta}/{2})\ket{1}\big]
   = e^{i\phi_{z^0}(t)}[ \cos({\theta}(t)/{2})\ket{0} +
   e^{i\phi(t)}\sin({\theta}(t)/{2})\ket{1}].\ee
It should be noted that in addition to the  $(\theta, \varphi)$ Bloch sphere characterization of the usual 2-state density matrix, the quantum {\it state} depends additionally on $e^{i\phi_{z^0}}$ which contains the geometric phase and connection.
\\
\\
{\bf A 2-state subsystem of the harmonic oscillator:}
\\
\\
A very simple example is the time-independent harmonic oscillator with Hamiltonian and eigenvalues,
$H=p^2/2m+\frac{1}{2}m\omega^2x^2$; \quad
$E_n=\left(n+\frac{1}{2}\right)\hbar\omega$. If we are restricted to a normalized 2-state basis $|n=0\rangle$ and $|n=1\rangle$, it follows that $|\Psi(0)\rangle=\cos(\theta/2)|0\rangle+\sin(\theta/2)|1\rangle$; and
\be|\Psi(t)\rangle=e^{-\frac{i}{\hbar}Ht}|\Psi(0)\rangle&=&\cos(\theta/2)e^{-\frac{i\omega
t}{2}}|0\rangle+\sin(\theta/2)e^{-\frac{3i\omega t}{2}}|1\rangle\cr
&=&e^{-\frac{i\omega
t}{2}}\left[\cos(\theta/2)|0\rangle+\sin(\theta/2)e^{-i\omega
t}|1\rangle\right].\ee
Thus we read off
$\phi_{z^0}=-\frac{\omega t}{2}$; $\phi=-\omega t$;
$A=-d\phi_{z^0}+d\chi-\frac{1}{2}\cos\theta d\phi.$
At time $t=T=2\pi/\omega$, we then have $|\Psi(t=2\pi/\omega)\rangle=e^{-i\pi}|\Psi(0)\rangle$. Furthermore,
$\int^{t=2\pi/\omega}_0 A=\int^{2\pi/\omega}_0
\frac{\omega}{2}dt-\omega dt+\frac{1}{2}(\cos\theta)\omega
dt=\pi(\cos\theta-1)$ and
$\frac{1}{\hbar}\int^{t=2\pi/\omega}_0\langle\Psi(t')|H|\Psi(t')\rangle
dt'=(2-\cos\theta)\pi$. It can also be verified explicitly that these results confirm the general formula
\be \langle\Psi(T)\ket{\Psi(o)} &=&
 \frac{{\bar\zeta}^\alpha(T)\zeta^\alpha(o)}
       {[{\bar\zeta}^\beta(T)\zeta^\beta(T)]^{\frac{1}{2}}
[{\bar\zeta}^\kappa(o)\zeta^\kappa(o)]^{\frac{1}{2}}}e^{-i({\phi}_{z^0}(T)-{\phi}_{z^0}(o))}\cr
&=&\frac{{\bar\zeta}^\alpha(T)\zeta^\alpha(o)}{[{\bar\zeta}^\beta(T)\zeta^\beta(T)]^{\frac{1}{2}}
[{\bar\zeta}^\kappa(o)\zeta^\kappa(o)]^{\frac{1}{2}}}
 \exp\left(i\int^{\zeta(T)}_{\zeta(o)}A +\frac{i}{\hbar}\int^T_o\,\langle{\Psi(t)}|H(t)\ket{\Psi(t)}dt\right).
\ee
\\
{\bf Arbitrary spin J system in a rotating magnetic field:}
\\
\\
Consider, as shown in the figure below, a particle of angular momentum J  in a rotating magnetic field
$\vec{B}=B(\sin\alpha\cos\omega t,\sin\alpha\sin\omega
t,\cos\alpha)$ inclined at angle $\alpha$  with respect to the z-axis.
\begin{figure}[h!]
\centering
\includegraphics[angle=-90,width=0.4\textwidth]{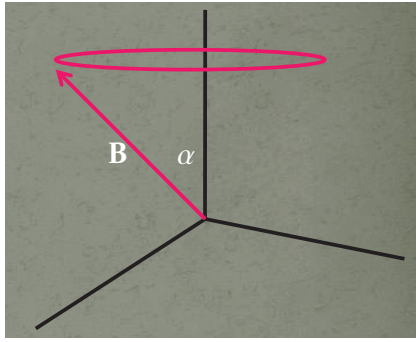}
\caption{The rotating magnetic field.}
\end{figure}\\
The time-dependent Hamiltonian of the system has the form $H(t)=-\mu B(\sin\alpha\cos\omega t
J_1+\sin\alpha\sin\omega t J_2+\cos\alpha
J_3).$ with $H(0)=-\mu B(\sin\alpha J_1
+\cos\alpha J_3)$. Furthermore, the Hamiltonian
\be H(t)=V^{\dag}H(0)V=e^{-i\omega tJ_3}[-\mu B(\sin\alpha J_1
+\cos\alpha J_3)]e^{i\omega tJ_3}; \quad V=e^{i\omega t J_3}\ee
does not commute at different times; and the evolution of the state is governed by
$|\Psi(t)\rangle=U(t)|\Psi(0)\rangle,  i\hbar\frac{d}{dt}|\Psi(t)\rangle=H(t)|\Psi(t)\rangle$ with time-ordered evolution operator,
$U(t) = T[\exp(-\frac{i}{\hbar}\int^T_{0} H(t') dt')]$. This implies,
\be H(0)=VH(t)V^{\dag}=V\left[i\hbar(\frac{d}{dt}U)U^{\dag}\right]V^{\dag}
=i\hbar(-i\omega J_3)+i\hbar(\frac{d}{dt}VU)(VU)^{\dag};\ee
with
$U(t)=V^{\dag}\exp\left\{i\Bigl[\frac{\mu B}{\hbar}\sin\alpha J_{1}+
(\frac{\mu B}{\hbar}\cos\alpha+\omega)J_{3}\Bigr]t\right\}=V^{\dag}\exp\left\{i[\Omega\tan\beta J_{1}+\Omega J_{3}]t\right\}.$
It can be worked out that the unitary evolution operator (for arbitrary J) takes the form
\be (U)_{MM'} &=&\left\langle J,
M|V^{\dag}\exp\left\{i\Bigl[\frac{\mu B}{\hbar}\sin\alpha J_{1}+
                          (\frac{\mu B}{\hbar}\cos\alpha+\omega)J_{3}\Bigr]\right\}
                       |J, M'\right\rangle \cr
&= &(e^{-\frac{i\omega
tJ_{z}}{\hbar}}e^{-\frac{iJ_{z}\gamma}{\hbar}}e^{-\frac{iJ_{y}\beta'}{\hbar}}
                                               e^{-\frac{iJ_{z}\alpha'}{\hbar}})_{MM'}\cr
&=& e^{-i[M\omega
t+(M+M')\gamma]}(\cos\frac{\beta'}{2})^{M+M'}(\sin\frac{\beta'}{2})^{M-M'}
             (-1)^{M-M'}e^{iM'\pi}\Bigl[\frac{(J-M)!(J-M')!}{(J+M)!(J+M')!}\Bigr]^{\frac{1}{2}}\cr
             &&\cdot\sum^{2J}_{n=0}(-1)^{n}
\frac{(J+M+n)!}{(J-M-n)!(M-M'+n)!\;n!}(\sin\frac{\beta'}{2})^{2n};
\ee
wherein
\be \sin\frac{\beta'}{2} &=&
\sin\beta\sin\frac{\vartheta t}{2}, \quad
 \sin\gamma = \frac{\cos\frac{\vartheta t}{2}}{\sqrt{\cos^2 \frac{\vartheta t}{2}+\cos^2 \beta\sin^2 \frac{\vartheta t}{2}}}, \quad
\cos\frac{\beta'}{2} = \sqrt{\cos^2 \frac{\vartheta t}{2}+\cos^2\beta\sin^2 \frac{\vartheta t}{2}},
\cr \cos\gamma &=& \frac{\cos\beta\sin\frac{\vartheta t}{2}}
                           {\sqrt{\cos^2 \frac{\vartheta t}{2}+\cos^2 \beta\sin^2 \frac{\vartheta t}{2}}},\quad
       \vartheta  =\frac{\Omega}{\cos\beta},
       \alpha' = \gamma - \pi.
\ee
\\
{\bf Qubit spin 1/2 system in rotating magnetic field:}
\\
\\
Specializing to a spin 1/2 or 2-state system in a rotating magnetic field, the evolution operator is then
\be U(t)= & \left(
  \begin{array}{cc}
    e^{-\frac{i\omega t}{2}}\left(\cos\frac{\vartheta t}{2}+i\cos\beta\sin\frac{\vartheta t}{2}\right)  &
    e^{\frac{-i\omega t}{2}}\left(i\sin\beta\sin\frac{\vartheta t}{2}\right)                           \\
    e^{\frac{i\omega t}{2}}\left(i\sin\beta\sin\frac{\vartheta t}{2}\right)                             &
    e^{\frac{i\omega t}{2}}\left(\cos\frac{\vartheta t}{2}-i\cos\beta\sin\frac{\vartheta t}{2}\right)  \\
  \end{array}
\right).
\ee
As an example, the choice of the initial state
$|\Psi(0)\rangle=\left(
                          \begin{array}{c}
                            1 \\
                            0 \\
                          \end{array}
                        \right),$ and
$|\Psi(t)\rangle=U(t)|\Psi(0)\rangle$ yield the following results:
\be
&&A=A_{(1)}=\frac{1}{4}\left(2\vartheta\cos\beta-\omega(3+\cos(2\beta)+2\cos(\vartheta t)\sin^2\beta)\right)dt,  \\
&&\frac{1}{\hbar}\langle\Psi(t)|H(t)|\Psi(t)\rangle=
\frac{1}{2}(-\vartheta \cos\beta+\omega\cos^2\beta+\omega\cos\vartheta t\sin^2\beta), \quad t\neq0 , \\
&&A+\frac{1}{\hbar}\langle\Psi(t)|H(t)|\Psi(t)\rangle
dt=-\frac{\omega}{2}dt;
\ee
\be\langle\Psi(T)|\Psi(t)\rangle=
&e^{-\frac{1}{2}i(t+T)\omega}
   \Big\{
     e^{it\omega}\sin^2\beta\sin\frac{\vartheta t}{2}\sin\frac{\vartheta T}{2}+
       e^{iT\omega}\Big(
                   \cos\frac{\vartheta t}{2}+i\cos\beta\sin\frac{\vartheta t}{2}
                   \Big)\nonumber \\
 &\cdot \Big(\cos\frac{\vartheta T}{2}-i\cos\beta\sin\frac{\vartheta T}{2}
                  \Big)
   \Big\}
\ee
\be
\frac{\bar{\zeta}^\alpha(T)\zeta^\alpha(t)}{[\bar{\zeta}^\beta(T)\zeta^\beta(T)]^{\frac{1}{2}}
                                            [\bar{\zeta}^\kappa(t)\zeta^\kappa(t)]^{\frac{1}{2}}}
= &&\sin^2\beta\cdot\sin\frac{\vartheta t}{2}\cdot\sin\frac{\vartheta T}{2}\nonumber\\
&+&e^{-i(t-T)\omega}
     \left(
      \cos\beta\sin\frac{\vartheta t}{2}-i\cos\frac{\vartheta t}{2}
     \right)
     \left(
      \cos\beta\sin\frac{\vartheta T}{2}+i\cos\frac{\vartheta T}{2}
     \right).
\ee
It can again be checked from these that the formula
$\langle\Psi(T)|\Psi(t)\rangle=\frac{
      \bar{\zeta}^\alpha(T)\zeta^\alpha(t)}{[\bar{\zeta}^\beta(T)\zeta^\beta(T)]^{\frac{1}{2}}
                                            [\bar{\zeta}^\kappa(t)\zeta^\kappa(t)]^{\frac{1}{2}}}
e^{
      i\int^T_t
         \left(
          A+\frac{1}{\hbar}\langle\Psi(t)|H(t)|\Psi(t)\rangle dt
         \right)}$ is satisfied.
\\
\\
{\bf Qutrit 3-state system:}
\\
\\
A 3-state  $J=1$ system placed rotating magnetic field has evolves in accordance with
\be U(t)=
 \left(
      \begin{array}{ccc}
        -\cos^2 \frac{\beta'}{2}e^{-i(2\gamma+\omega t)}    &
        -\frac{1}{\sqrt2}\sin\beta'e^{-i(\gamma+\omega t)}  &
        -\sin^2 \frac{\beta'}{2}e^{-i\omega t}             \\[5pt]
        -\frac{1}{\sqrt2}\sin\beta'e^{-i\gamma}             &
        \cos\beta'                                          &
        \frac{1}{\sqrt2}\sin\beta'e^{i\gamma}              \\[5pt]
        -\sin^2 \frac{\beta'}{2}e^{i\omega t}               &
        \frac{1}{\sqrt2}\sin\beta'e^{i(\gamma+\omega t)}    &
        -\cos^2 \frac{\beta'}{2}e^{i(2\gamma+\omega t)}
      \end{array}
   \right).\ee
\\
For arbitrary initial state to return to original value modulo an overall
phase (i.e. closed path in $CP^2$) after time T, the conditions
\be\omega T=2\pi; \quad \vartheta T=2m\pi \quad m=1, 2, 3, \cdots\ee must be satisfied.
We consider some specific cases as illustrations:
\\
{\bf Case (1):} Let $\frac{\hbar \omega}{\mu
B}=\frac{1}{\sqrt{26-5\sqrt2}}$, and $\frac{\hbar \vartheta}{\mu
B}=\frac{5}{\sqrt{26-5\sqrt2}}$. In this case, $\tan\beta$ is chosen
to be $1$, and $\vartheta=5\omega$. Thus $m=1$ and $n=5$ and the
periodicity conditions are satisfied. The expectation value
$\langle\Psi(t)|\overrightarrow{J}\ket{\Psi(t)}$ with initial state
$\langle m=1,0,
-1|\Psi(0)\rangle=(0,\frac{1}{\sqrt{2}},\frac{1}{\sqrt{2}})$ is
plotted in Fig.(2). Since in this case the final state differs from
the initial state by an overall phase, $\langle\Psi(t)|\overrightarrow{J}\ket{\Psi(t)}$ must be closed. And
it is.
\\
{\bf Cases (2) and (3):} We next choose $\frac{\hbar \omega}{\mu
B}=\frac{1}{\sqrt{\frac{7}{2}-\sqrt5}}$ and $\frac{\hbar
\vartheta}{\mu B}=\sqrt{\frac{5}{2(\frac{7}{2}-\sqrt5)}}$. The
irrational ratio for $\frac{\vartheta}{\omega}$ implies that states
with this configuration do not obey the periodicity conditions. For
the same initial state as in {\bf Case (1)}, the evolution for two
different final times $T$ are plotted in Fig.(3) and Fig.(4). As
expected, $\langle\Psi(t)|\vec{J}\ket{\Psi(t)}$ does not describe
closed paths in either instance, implying that the state cannot
differ by just an overall phase after time $T$ when the periodicity
conditions are not satisfied.
\begin{figure}[h!]
\centering
\includegraphics[width=0.3\textwidth]{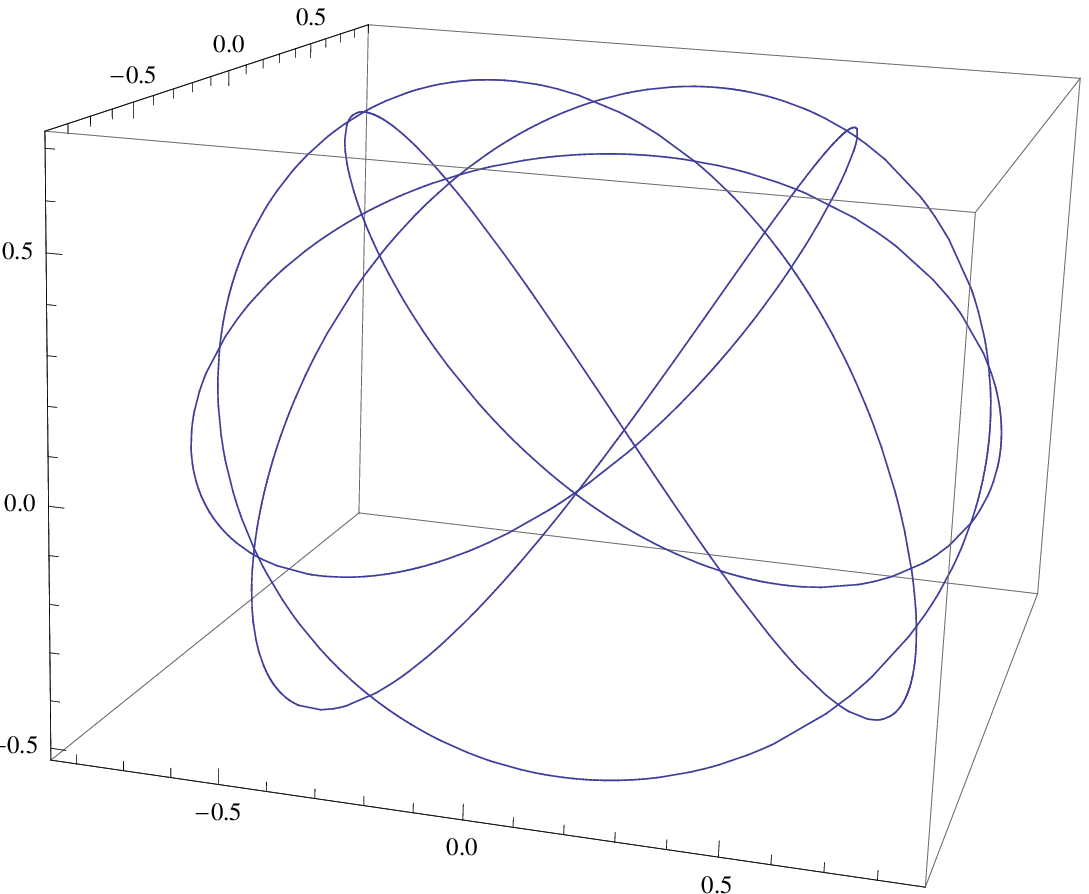}
\caption{$\langle\Psi(t)|\protect\vec{J}\ket{\Psi(t)}$ from $t=0$ to
$t=\frac{2\pi}{\omega}$ with $\vartheta=5\omega$.}
%\end{figure}
%\begin{figure}[h!]
\includegraphics[width=0.3\textwidth]{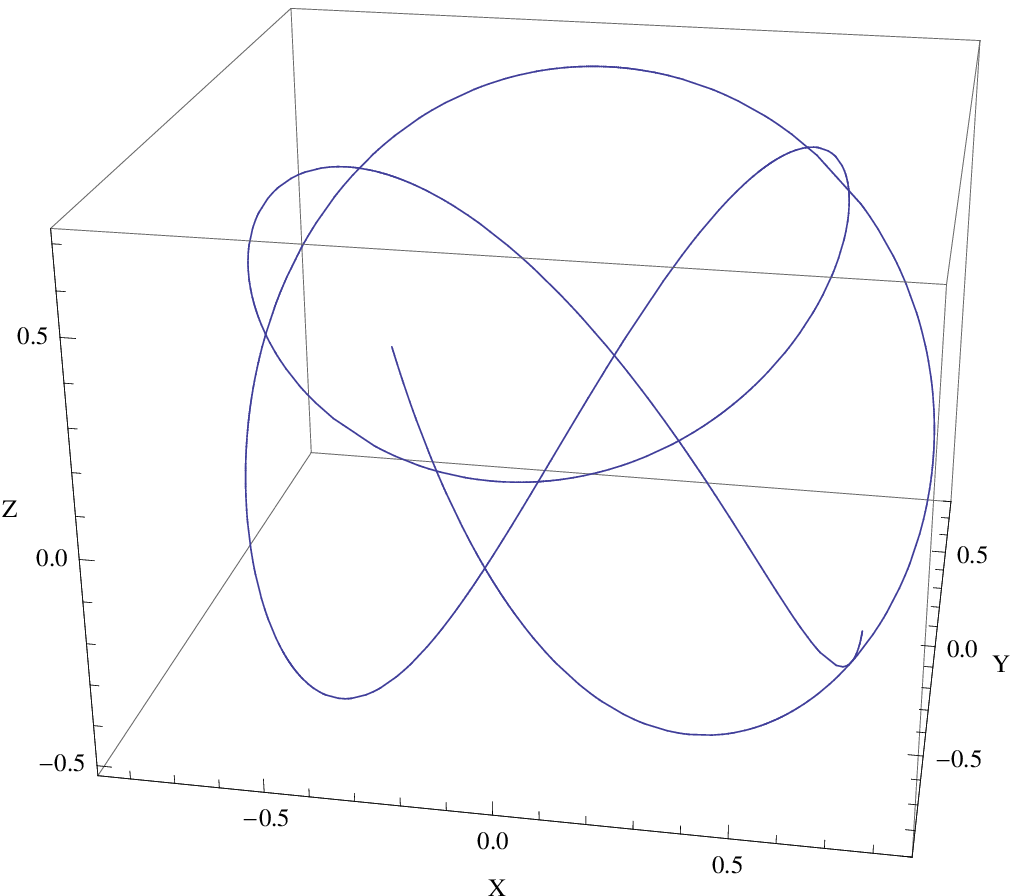}
\caption{$\langle\Psi(t)|\protect\vec{J}\ket{\Psi(t)}$ from $t=0$ to
$t=\frac{2\sqrt2\pi}{\omega}$.}
\includegraphics[width=0.3\textwidth]{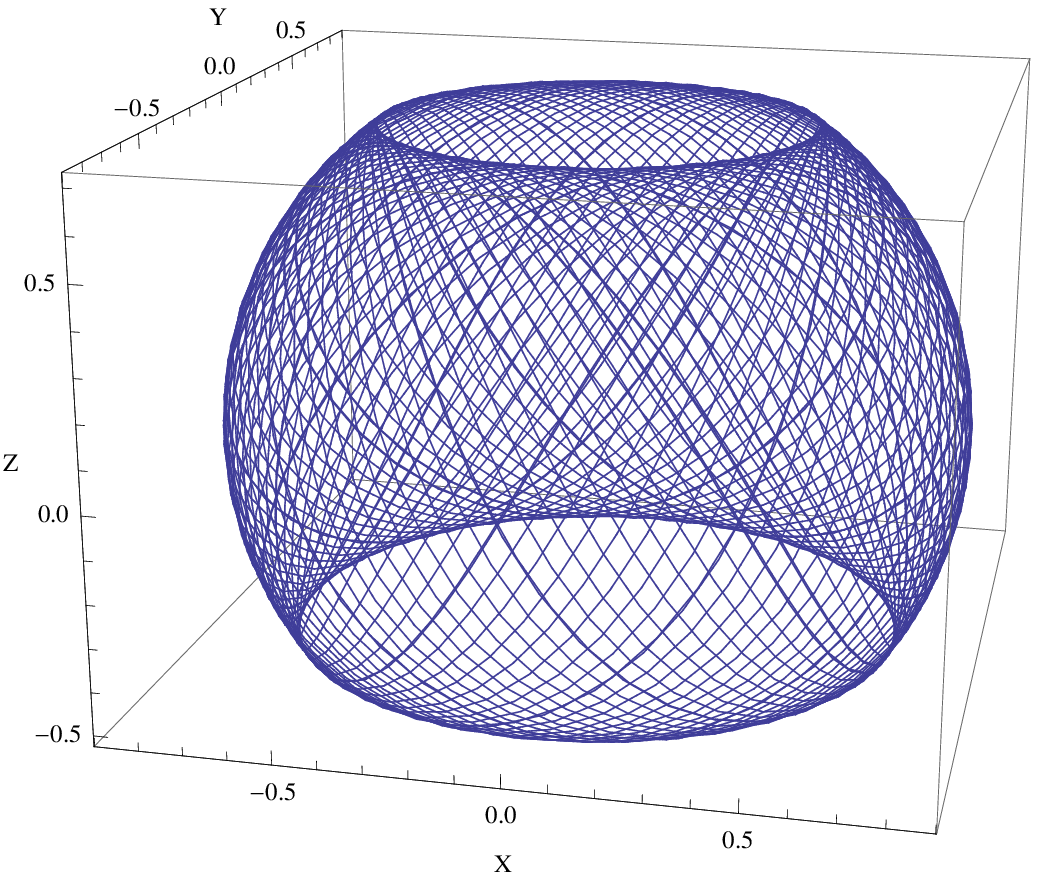}
\caption{$\langle\Psi(t)|\protect\vec{J}\ket{\Psi(t)}$ from $t=0$ to
$t=30\cdot\frac{2\pi}{\omega}$.}
\end{figure}
\\
\\
{\bf Qutrit 3-state system and $S^5/S^1=CP^2$ fibration:} It should be mentiond that a generic 3-state system corresponds to the
Hopf fibration $S^5/S^1=CP^2$. The state can be expressed as $|\Psi\rangle=c^0|0\rangle+c^1|1\rangle+c^2|2\rangle$, and
explicit parametrization of $S^5$  by
$ c^0=e^{i(\chi+\phi)}\cos(\theta_1/2),\quad c^1=e^{i(\chi-\phi)}\sin(\theta_1/2)\cos(\theta_2/2), \quad
c^2=e^{i(\phi_3)}\sin(\theta_1/2)\sin(\theta_2/2),$
lead to
\be \zeta^0_{(0)}=1, \quad
\zeta^1_{(0)}=e^{2i\phi}\tan(\theta_1/2)\cos(\theta_2/2),\quad
\zeta^2_{(0)}=e^{i\gamma}\tan(\theta_1/2)\sin(\theta_2/2), \gamma\equiv\phi_3-\chi+\phi. \ee
From these, it can be computed that
$A=\frac{1}{4}(1-\cos\theta_1)\left[d(2\phi+\gamma)+\cos\theta_2d(2\phi-\gamma)\right]$, and its curvature 2-form
\be F=\frac{1}{2\pi}\sin\theta_1d\theta_1\wedge\left[2\cos^2(\theta_2/2)d\phi+\sin^2(\theta_2/2)d\gamma\right]
-(1-\cos\theta_1)\sin\theta_2d\theta_2\wedge d(2\phi-\gamma)=\ast
F.\ee
Moreover, it can be checked that integrated over the entire 4-dimensional $CP^2$ manifold the self-dual curvature $F$ yields
$\frac{1}{4\pi^2}\int_{CP^2}F\wedge F=+1$.

\stepcounter{section}
\section*{Quaternionic Hopf fibration and non-Abelian geometric phase factor}
In a manner analogous to the construction of Hopf fibration over complex projective space, the formalism for quaternionic projective space can be obtained by studying the geometry of $S^{4K+3}/[S^3=SU(2)] = HP^K$, but with the caveat that it is applicable only to finite-dimensional systems with {\it even} number of states. The reason is that each quaternion has to be associated with a pair of complex state coefficients. Starting with a generic state as
\be |\Psi\rangle=\sum\limits^{N=2K+1}_{a=0}C^a|a\rangle; \qquad C^a=\sum\limits^N_{b=a}\frac{z^a}{\sqrt{z^b\bar{z}^b}};\ee
we may defined the associated quaternions through
\be q^\alpha=Re(z^\alpha)I_2+Im(z^\alpha)\frac{\sigma^1}{i}+Re(z^{\underline{\alpha}})\frac{\sigma^2}{i}
+Im(z^{\underline{\alpha}})\frac{\sigma^3}{i};  \quad \alpha=0, 1,
\cdots, K, \underline{\alpha}\equiv\alpha+K+1\quad \sigma^{1,2,3} ={\rm Pauli\, matrices}.\ee
It follows that
\be |\Psi\rangle=\sum\limits^K_{\alpha=0}Tr\left(P^{-}_1Q^\alpha\right)|\alpha\rangle+Tr\left(P^{+}_1(i\sigma^2)Q^\alpha\right),
\quad Q^\alpha\equiv\frac{q^\alpha}{\sqrt{\frac{1}{2}Tr\left(q^\beta
q^{\dag\beta}\right)}};\ee
with the projector $P^{\pm}=\frac{1}{2}(I\pm\sigma^1)$. For the bundle $S^{4K+3}/S^3=HP^K$,  the quaternionic Hopf fibration can be realized through the projection
$H^{K+1}-\left\{0\right\} \rightarrow S^{4K+3} \rightarrow HP^K$  which is
\be q^\alpha \rightarrow
Q^\alpha=\frac{q^\alpha}{\sqrt{q^\beta q^{\dag\beta}}}
\rightarrow Q^\alpha/Q^\eta=q^\alpha/q^\eta=h^\alpha_{(\eta)},
\qquad \sum\limits_\alpha\left|Q^\alpha\right|^2=1; \ee
in each local chart $U^{(\eta)}$ and extended to the atlas $\cup U^{(\eta)}$. The resultant $h^\alpha\in HP^K$ are precisely the inhomogeneous coordinates of $HP^K$. In $U^{(0)}$,  $h^\alpha\equiv(q^0)^{-1}q^\alpha=(Q^0)^{-1}Q^\alpha, \quad \forall
0\leq\alpha\leq K$. It follows that the expression for the state expressed in terms of local $HP^K$ coordinates and the fiber (which is the
unit quaternion $\hat{q}^0\equiv q^0/\left|q^0\right|\in SU(2)=S^3$) is thus
\be \left|\Psi\right>=\sum\limits^K_{\alpha,\beta=0}\frac{Tr(\mathsf{P}^-_1\hat{q}^0h^\alpha)\left|\alpha\right>+Tr(\mathsf{P}^+_1(i\sigma^2)\hat{q}^0h^\alpha)\left|\underline{\alpha}\right>}{\sqrt{\frac{1}{2}Tr(h^\beta
h^{\dagger\beta})}}.\ee
By substituting into the Schrodinger equation, we can similarly obtain the evolution of the non-Abelian phase factor
\be \hat{q}^0(t)=\mathcal{T}e^{\frac{i}{\hbar}\int^t_0\mathcal{H}dt}\hat{q}^0(0)[\mathcal{T}e^{-i\int^t_0 A dt}]^\dagger;
\ee
wherein
$\mathcal{A}=Adt=\frac{h^\alpha(dh^{\dagger\alpha})/(dt)-(dh^\alpha)/(dt)h^{\dagger\alpha}}{iTr(h^\beta
h^{\dagger\beta})}dt$ is the associated non-Abelian connection. The connection $-\mathcal {A} =A_{HP^K}=\frac{dh^\alpha h^{\dag\alpha}-h^\alpha dh^{\dag\alpha}}{iTr(h^\beta h^{\dag\beta})}$ is also the (quaternionic) Kahler connection of the quaternionic Kahler manifold $HP^K$.  The overlap function at different times can be computed to be (analogous to the result of Eq. (9))
\be \left<\Psi(T)|\Psi(0)\right>=Tr\left(\frac{h^{\dagger\alpha}(T)}{\sqrt{\frac{1}{2}Tr(h^\beta(T)h^{\dagger\beta}(T))}}\mathcal{T}e^{-i\int^T_0Adt}\hat{q}^{\dagger0}(0)[\mathcal{T}e^{\frac{i}{\hbar}\int^T_0\mathcal{H}dt}]^\dagger\mathsf{P}^-_1\hat{q}^0(0)\frac{h^\alpha(0)}{\sqrt{\frac{1}{2}Tr(h^\gamma(0)
h^{\dagger\gamma}(0))}}\right); \ee
wherein \be
\frac{i}{\hbar}\mathcal{H}\equiv \frac{i}{\hbar}\left(\begin{array}{cc}
Re\left<\Psi^\bot|H|\Psi\right> & \left<\Psi|H|\Psi\right>+iIm\left<\Psi^\bot|H|\Psi\right>\\
\left<\Psi|H|\Psi\right>-iIm\left<\Psi^\bot|H|\Psi\right> &
-Re\left<\Psi^\bot|H|\Psi\right>\end{array}\right),\nonumber\ee and  $ \left|\Psi^\bot\right> =\sum\limits^K_{\alpha=0}Tr(\mathsf{P}^-_1\frac{\sigma^2}{i}Q^\alpha)\left|\alpha\right>
+Tr(\mathsf{P}^+_1Q^\alpha)\left|\underline{\alpha}\right>$. This is the complete and exact result revealing the $SU(2)$ geometric phase factor of arbitrary finite even-dimensional pure systems.

\subsection*{Explicit Example: Generic Four-State Systems and the BPST instanton}
A generic pure 4-state system is associated with the quaternionic Hopf fibration, $S^7/SU(2) =HP^1 =S^4$ .
Parametrization of $S^7$ can be achieved in terms of two quaternions $S^7 \sim (Q^0,Q^1)$ satisfying $\left|Q^0\right|^2+\left|Q^1\right|^2=1$. These can in turn be explicitly written as $Q^0=u\cos\frac{\theta}{2}$, $Q^1=uv\sin\frac{\theta}{2}$, with
\be u=\left(\begin{array}{cc}
e^{i(\gamma_1+\beta_1)/2}\cos{\frac{\alpha_1}{2}} &
e^{i(\gamma_1-\beta_1)/2}\sin{\frac{\alpha_1}{2}}\\
-e^{-i(\gamma_1-\beta_1)/2}\sin{\frac{\alpha_1}{2}} &
e^{-i(\gamma_1+\beta_1)/2}\cos{\frac{\alpha_1}{2}}\end{array}\right),
\quad v=\left(\begin{array}{cc}
e^{i(\gamma_2+\beta_2)/2}\cos{\frac{\alpha_2}{2}} &
e^{i(\gamma_2-\beta_2)/2}\sin{\frac{\alpha_2}{2}}\\
-e^{-i(\gamma_2-\beta_2)/2}\sin{\frac{\alpha_2}{2}} &
e^{-i(\gamma_2+\beta_2)/2}\cos{\frac{\alpha_2}{2}}\end{array}\right)\ee
being $SU(2)$ matrices. This yields,
\be\left|\Psi\right>=\sum\limits^1_{\alpha,\beta=0}\frac{Tr(\mathsf{P}^-_1\hat{q}^0h^\alpha)\left|\alpha\right>+Tr(\mathsf{P}^+_1(i\sigma^2)\hat{q}^0h^\alpha)\left|\underline{\alpha}\right>}{\sqrt{\frac{1}{2}Tr(h^\beta
h^{\dagger\beta})}},\ee and the non-Abelian connection corresponds exactly to the
BPST instanton\cite{BPST} gauge potential
$\mathcal{A}_{HP^1}=\frac{dh^\alpha
h^{\dag\alpha}-h^\alpha dh^{\dag\alpha}}{iTr(h^\beta h^{\dag\beta})}=-i\sin^2\frac{\theta}{2}dvv^\dag$. The  parameter $\theta$ is related to the distance $|x|$ from the instanton center by $\sin^2\left(\frac{\theta}{2}\right)=\left|x\right|^2/\left(\left|x\right|^2+\Lambda^2\right)$. The second Chern number is
computed to be $\mathcal{C}_2=-\frac{1}{8\pi^2}\int_{S^4}Tr(\mathcal{F}\wedge\mathcal{F})
=\frac{1}{8\pi^2}\int_{S^4}\frac{\sin^3\theta}{4}d\theta\wedge
Tr(dvv^\dag)^3 =\frac{1}{24\pi^2}\int_{S^3}Tr(dvv^\dagger)^3=1. $

\subsection*{Pure state bipartite qubit-qubit entanglement and the BPST instanton}
A bipartite qubit-qubit system is an example of a composite 4-state system.
In general the qubit-qubit system may be written as $|\Psi\rangle=c_{ij}|i\rangle|j\rangle$, where $i, j$ takes value
$\pm$. A quantitative measure of the pure qubit-qubit state entanglement is provided by the expectation value of the
Clauser-Horne-Shimony-Holt operator\cite{CHSH} which is
\be CHSH=(R+S)\otimes T+(R-S)\otimes U; \ee
wherein $R=\hat{\vec{r}}\cdot\sigma$ with $\hat{\vec{r}}$ being a unit spatial vector, and similarly the operators $S$, $T$ and $U$.
The expectation value of the CHSH operator depends on the state and also the directions of the unit vectors; but the maximum value\cite{KS} is correlated to the entanglement by $\langle\Psi|CHSH|\Psi\rangle_{max.}=2\sqrt{1+4\left|\det
c\right|^2}$, with $0\leq\left|\det c\right|^2\leq\frac{1}{4}$; wherein $\det c$ denotes the determinant of the $2\times 2$ state coefficient matrix $c_{ij}$.
Comparing with our generic 4-state system,
\be \left|\Psi(t)\right>=\sum\limits^{3}_{a=0}C^a|a\rangle= \sum\limits^1_{\alpha,\beta=0}\frac{Tr(\mathsf{P}^-_1\hat{q}^0(t)h^\alpha(t))\left|\alpha\right>
+Tr(\mathsf{P}^+_1(i\sigma^2)\hat{q}^0(t)h^\alpha(t))\left|\underline{\alpha}\right>}{\sqrt{\frac{1}{2}Tr(h^\beta(t)
h^{\dagger\beta}(t))}}, \qquad Q^0=u\cos\frac{\theta}{2}, \quad Q^1=uv\sin\frac{\theta}{2};\ee
and bearing in mind $q^\alpha=Re(z^\alpha)I_2+Im(z^\alpha)\frac{\sigma^1}{i}+Re(z^{\underline{\alpha}})\frac{\sigma^2}{i}
+Im(z^{\underline{\alpha}})\frac{\sigma^3}{i}$, the four state coefficients $C^a$ of the composite system can be computed to be
\begin{eqnarray*}
C^0=\cos\frac{\theta}{2}\cos\frac{\alpha_1}{2}e^{i(\gamma_1+\beta_1)/2}\quad,&\quad
C^1=\sin\frac{\theta}{2}\left(e^{i(\gamma_1+\gamma_2+\beta_1+\beta_2)/2}\cos\frac{\alpha_1}{2}\cos\frac{\alpha_2}{2}-
e^{i(\gamma_1-\gamma_2-\beta_1+\beta_2)/2}\sin\frac{\alpha_1}{2}\sin\frac{\alpha_2}{2}\right)\\
C^2=\cos\frac{\theta}{2}\sin\frac{\alpha_1}{2}e^{i(\gamma_1-\beta_1)/2}\quad,&\quad
C^3=\sin\frac{\theta}{2}\left(e^{i(\gamma_1+\gamma_2+\beta_1-\beta_2)/2}\cos\frac{\alpha_1}{2}\sin\frac{\alpha_2}{2}+
e^{i(\gamma_1-\gamma_2-\beta_1-\beta_2)/2}\sin\frac{\alpha_1}{2}\cos\frac{\alpha_2}{2}\right)
\end{eqnarray*}
A judicious mapping of the bipartite system to the composite system allows us to correlate the entanglement directly to the instanton parameter $\theta$ for arbitrary qubit-qubit 4-state composite systems. An earlier correlation can be found in \cite{Buras}.
The general unitary basis transformation between the two systems, $|\Psi\rangle=c_{ij}|i\rangle\otimes|j\rangle=C^a|a\rangle$, is $c_{ij}=\langle ij|a\rangle C^a=U^a_{ij}C^a$,
wherein
\be Tr\left(U^aU^{\dag b}\right)=U^a_{ij}(U^{\dag
b})_{ji}=U^a_{ij}(U^b_{ij})^*=\langle ij|a\rangle\langle
b|ij\rangle=\langle ij|b\rangle\langle a|ij\rangle=\delta^{ab}.\ee
To wit
\be \det c&=&\frac{1}{2}\epsilon^{ij}\epsilon^{kl}c_{ik}c_{jl}=\frac{1}{2}(i\sigma_2)^{ij}(i\sigma_2)^{kl}c_{ik}c_{jl}
=\frac{1}{2}Tr(\sigma^2c^T\sigma^2c)\\ \nonumber
&=&\frac{1}{2}Tr(\sigma^2(U^aC^a)^T\sigma^2(U^bC^b))
=\frac{1}{2}Tr(\sigma^2(\widetilde{U}^a)^T\sigma^2\widetilde{U}^b)|C^a||C^b|;
\ee
wherein we have defined $\widetilde{U}^a \equiv e^{i\phi_a}U^a$ satisfying $Tr(\widetilde{U}^a\widetilde{U}^{\dagger
b})=e^{i(\phi_a-\phi_b)}Tr(U^aU^{\dagger
b})=e^{i(\phi_a-\phi_b)}\delta^{ab}=\delta^{ab}$. The choice of $\tilde{U}^a=u^a\frac{\sigma^a}{\sqrt2}$ (no sum over $a$)
with $u^0=i$, $u^1=1$, $u^2=i$, $u^3=1$ yields the desired result
\be
\det c&=&-\frac{1}{4}Tr((\sigma^a)^\dagger\sigma^b)u^au^b|C^a||C^b|
=-\frac{1}{2}((u^0)^2|C^0|^2+(u^1)^2|C^1|^2+(u^2)^2|C^2|^2+(u^3)^2|C^3|^2)\cr
&=&\frac{1}{2}(|C^0|^2+|C^2|^2-|C^1|^2-|C^3|^2)=\frac{1}{2}(\cos^2\frac{\theta}{2}-\sin^2\frac{\theta}{2})
=\frac{1}{2}\cos\theta.
\ee
This relates the entanglement parameter, $\det c$, to the instanton parameter, $\theta$, in a generic qubit-qubit system which is considered as a 4-state total system with the Hilbert space $S^7$ which is total bundle space of the Hopf fibration $S^7/[SU(2)=S^3] = [HP^1=S^4]$.
\\
\stepcounter{section}
\section*{Summary}\small
The existence of Hopf fibrations $S^{2N+1}/S^1 = CP^N$ and $S^{4K+3}/S^3 = HP^K$ allows us to treat the
Hilbert space of generic finite-dimensional quantum systems as the total bundle space with
respectively $U(1)$ and $SU(2)$ fibers and complex and quaternionic projective  base manifolds.
This alternative method of studying and describing quantum states and their evolution reveals the intimate and exact connection between generic quantum systems and fundamental geometrical objects.

\stepcounter{section}
\subsection*{Acknowledgement}\small
The research for this work is supported in part by the National Science Council of Taiwan under
Grant No. NSC101-2112-M-006-007-MY3 and the National Center for Theoretical Sciences.
CS would like to thank C. H. Oh and K. Singh for beneficial discussions at the Centre for Quantum Technologies, Nat. U. Singapore, where part of this work was completed.

\end{document}